\documentclass[article]{llncs}
 \usepackage{xcolor}
\usepackage{caption} 
 \usepackage{xcolor}
\usepackage{amsmath}
\usepackage{graphicx}
\usepackage{courier}
\usepackage{caption}
\usepackage{xcolor}
\definecolorset{rgb/hsb/cmyk/gray}{}{}%
 {red,1,0,0/0,1,1/0,1,1,0/.3;%
  green,0,1,0/.33333,1,1/1,0,1,0/.59;%
  blue,0,0,1/.66667,1,1/1,1,0,0/.11;%
  brown,.75,.5,.25/.083333,.66667,.75/0,.25,.5,.25/.5475;%
  lime,.75,1,0/.20833,1,1/.25,0,1,0/.815;%
  orange,1,.5,0/.083333,1,1/0,.5,1,0/.595;%
  pink,1,.75,.75/0,.25,1/0,.25,.25,0/.825;%
  purple,.75,0,.25/.94444,1,.75/0,.75,.5,.25/.2525;%
  teal,0,.5,.5/.5,1,.5/.5,0,0,.5/.35;%
  violet,.5,0,.5/.83333,1,.5/0,.5,0,.5/.205}%
\definecolorset{cmyk/rgb/hsb/gray}{}{}%
 {cyan,1,0,0,0/0,1,1/.5,1,1/.7;%
  magenta,0,1,0,0/1,0,1/.83333,1,1/.41;%
  yellow,0,0,1,0/1,1,0/.16667,1,1/.89;%
  olive,0,0,1,.5/.5,.5,0/.16667,1,.5/.39}
\definecolorset{gray/rgb/hsb/cmyk}{}{}%
 {black,0/0,0,0/0,0,0/0,0,0,1;%
  darkgray,.25/.25,.25,.25/0,0,.25/0,0,0,.75;%
  gray,.5/.5,.5,.5/0,0,.5/0,0,0,.5;%
  lightgray,.75/.75,.75,.75/0,0,.75/0,0,0,.25;%
  white,1/1,1,1/0,0,1/0,0,0,0}

\DeclareCaptionFont{white}{\color{white}}
\DeclareCaptionFormat{listing}{\colorbox{gray}{\parbox{\textwidth}{#1#2#3}}}
\newfont{\code}{cmss10}

\newfont{\codeF}{cmss8}

\usepackage{color}
\usepackage{multirow}
\usepackage[latin1]{inputenc}
\usepackage[T1]{fontenc}
\usepackage[english]{babel}
\usepackage[normalem]{ulem}
\usepackage{verbatim}
\usepackage{listings}
\usepackage{graphicx}
\usepackage{cite}
\usepackage{indentfirst}
\usepackage{algorithmic,algorithm,pseudocode}
\usepackage{amssymb,amsfonts,amsxtra}
\usepackage[graphicx]{realboxes}
\usepackage{multicol}

\title{Leveraging Data Preparation, \\ HBase NoSQL Storage, and HiveQL Querying for COVID-19 Big Data Analytics Projects \\ \vspace{1cm} Version 1.0 - {Marsh 31, 2020}}

\author{Karim Ba\"{\i}na}
\institute{
Alqualsadi research team (Innovation on Digital and Enterprise Architectures) \\
ADMIR Laboratory, Rabat IT Center, \\
 ENSIAS, University Mohammed V in Rabat, \\
 BP 713 Agdal, Rabat, Morocco\\
\email{karim.baina@um5.ac.ma}}

\usepackage{url}

\begin{document}


 \maketitle \smallskip {\rmfamily
\begin{abstract}
Epidemiologist, Scientists, Statisticians, Historians, Data engineers and Data scientists are working on finding descriptive models and theories to explain COVID-19 expansion phenomena or on building  analytics predictive models for learning the apex of COVID-19 confimed cases, recovered cases, and deaths evolution curves. In CRISP-DM life cycle, 75\% of time is consumed only by data preparation phase causing lot of pressions and stress on scientists and data scientists building machine learning models. This paper aims to help reducing data preparation efforts by presenting detailed schemas design and data preparation technical scripts for formatting and storing Johns Hopkins University COVID-19 daily data in {\codeF HBase} NoSQL data store, and enabling {\codeF HiveQL} COVID-19 data querying in a relational {\codeF Hive} SQL-like style.
\end{abstract}

\newcommand{\algvar}[1]{\text{\ttfamily\upshape#1}}

\definecolor{codegreen}{rgb}{0,0.6,0}
\definecolor{codegray}{rgb}{0.5,0.5,0.5}
\definecolor{codepurple}{rgb}{0.58,0,0.82}
\definecolor{backcolour}{rgb}{0.95,0.95,0.92}

\lstdefinestyle{mystyle}{
    backgroundcolor=\color{backcolour},   
    commentstyle=\color{codegreen},
    keywordstyle=\color{magenta},
    numberstyle=\tiny\color{codegray},
    stringstyle=\color{codepurple},
    basicstyle=\ttfamily\footnotesize,
    breakatwhitespace=false,         
    breaklines=true,                 
    captionpos=b,                    
    keepspaces=true,                 
    numbers=left,                    
    numbersep=5pt,                  
    showspaces=false,                
    showstringspaces=false,
    showtabs=false,                  
    tabsize=2
}

\lstdefinestyle{tt}{basicstyle=\small\ttfamily,keywordstyle=\bfseries,language=[LaTeX]{TeX}}
\lstdefinestyle{rm}{basicstyle=\ttfamily,keywordstyle=\slshape,language=[LaTeX]{TeX}}

\lstdefinestyle{numberstyle1}{
 basicstyle=\footnotesize\ttfamily, 
    numbers=left,              
    numberstyle=\tiny,          
    numbersep=5pt,              
    tabsize=2,                  
    extendedchars=true,
    breaklines=true,            
    keywordstyle=\color{red},
    frame=b,
    stringstyle=\color{white}\ttfamily, 
    showspaces=false,
    showtabs=false,
    xleftmargin=17pt,
    framexleftmargin=17pt,
    framexrightmargin=5pt,
    framexbottommargin=4pt,
    showstringspaces=false
}
\lstset{style=numberstyle1}

\section{Introduction}

Johns Hopkins University has provided a github repository with, among others, daily fresh data about COVID-19 pandemy confirmed cases, recovered cases, and deaths evolution \cite{jhu2020}. Epidemiologist, Scientists, Statisticians, Historians, Data engineers and Data scientists are working on finding models and theories to explain and predict COVID-19 expansion phenomena. Our paper aims to help reducing data preparation efforts by presenting detailed schemas design and data preparation technical scripts for formatting and storing Johns Hopkins University COVID-19 daily data in {\code HBase}\cite{hbase2020} NoSQL data store, and enabling {\code HiveQL} COVID-19 data querying in a relational {\code Hive}\cite{hive2020} SQL-like style. Our data integration and analytics approach for this paper, and for handling COVID-19 crisis in general is building  Minimum Viable Model, Platform, and Data Product through agile analytics \cite{tan2018}.

\section{Prerequisites}
(1) {\code Hadoop} or {\code YARN} \cite{hadoop2020} with {\code HBase}, {\code Hive} need to be installed\footnote{In this paper, a minimal IBM BigInsights quick start VM is used as a DEV on-premises Big Data platform, however, scripts are compatible with every on-premises Hadoop Apache compliant Big Data distributions (like Cloudera/Hortonworks HDP, IBM Open Platform, MapR, etc.), or Hadoop Apache compliant Cloud Analytics Solutions (like Google Cloud Big Data Analytics Solutions, Microsoft Azure HDInsight, Amazon AWS EMR, IBM Analytics Engine on IBM Watson, etc.)}. (3) A Linux/Unix environment is requeted for running the given shell scripts. (4) {\code git} tools are necessary for pulling data.


\begin{minipage}{\textwidth}
\begin{lstlisting}[caption=Targeted Datalake Hadoop ecosystem starting]
./start.sh hadoop
./start.sh hbase
./start.sh hive

\end{lstlisting}
\end{minipage}

This step depends on the {\code Hadoop/YARN} environment specificities. {\code Ambari}\cite{ambari2020} may be used to start those services.

\section{COVID-19 Data Preparation}

\subsection{Data Collection}

COVID-19 Data can be collected each day from Johns Hopkins University Center for Systems Science and Engineering (JHU CCSE) github repository around 2.00 am GMT+1.

\begin{minipage}{\textwidth}
\begin{lstlisting}[caption=Pulling COVID-19 Data]
#kbaina is my local home directory, change it to your home directory or another directory
cd /home/kbaina/

git clone https://github.com/CSSEGISandData/COVID-19.git COVID-19/

cd ./COVID-19/

git pull
\end{lstlisting}
\end{minipage}

\subsection{Data Formatting}
{\code ingest}\_{\code and}\_{\code clean.sh} data fomatting Shell script removes $\backslash"$, '$*$' characters, and replaces non separator ',' by '-' character (e.g. in "Korea, South"), formats column dates into "\%m/\%d/\%Y" format (eg. 3/2/20 becomes 03/02/2020) enabling dates operations, keeps only not null values from the sparse matrix, and merges the two first columns to form a composite key separated by a '{\textasciitilde{}}' character.

\begin{minipage}{\textwidth}
\begin{lstlisting}[caption=Data Fomatting Shell Script ({\code ingest}\_{\code and}\_{\code clean.sh})]
#!/bin/sh
specific=$1

#$1 script parameter may be 'confimed' or 'deaths' or 'recovered' 

sed "s/, /-/" ./COVID-19/csse_covid_19_data/csse_covid_19_time_series/
time_series_covid19_${specific}_global.csv | sed "s/\"//g" |
sed "s/\*//" | sed -E "s/\,(.)\//,0\1\//g" |
sed -E "s/\/(.)\//\/0\1\//g" |
sed -E "s/\/20([^/])/\/2020\1/g" |
sed -E "s/\/20$/\/2020/g"| sed -E "s/,($)/,0\1/g" |
sed "s/,0,/,,/g"| sed  -E "s/([,]+)0,/\1,/g" |
sed "s/,0$/,$/" | sed  "s/^,/~/" |
sed -E "s/([a-z A-Z]+),([a-z A-Z]+)/\1~\2/" > 
time_series_covid19_${specific}_global-sparse-with-formatted-column-names.csv

# time_series_covid19_${specific}_global-sparse-with-formatted-column-names.csv
# contains date formatted columns useful for any further date arithmetics and manipulation
# (e.g. duration calculations, D0 of COVID-19, D0 of n^th death manipulation, etc.)

tail -n +2 time_series_covid19_${specific}_global-sparse-with-formatted-column-names.csv > time_series_covid19_${specific}_global-sparse.csv
\end{lstlisting}
\end{minipage}

\begin{minipage}{\textwidth}
\begin{lstlisting}[caption=Calling Data Fomatting Shell Script ({\code ingest}\_{\code and}\_{\code clean.sh})]
cd /home/kbaina/

chmod u+x ./ingest_and_clean.sh

./ingest_and_clean.sh confirmed

./ingest_and_clean.sh deaths

#next command will succeed but you should adapt next section scripts for creating, feeding, and querying recovered tables.
#./ingest_and_clean.sh recovered

\end{lstlisting}
\end{minipage}

\section{NoSQL HBase Storage and Hive SQL/pure NoSQL interoperability}
In this section present NoSQL and relational schema design and detailed technical scripts for storing  JHU COVID-19 daily confimed cases and deaths data \footnote{without loosing in generality, all scripts in this paper can be very easily adapted to take into account JHU CCSE recovered cases file ({\codeF time}\_{\codeF series}\_{\codeF  covid19}\_{\codeF recovered}\_{\codeF global}.{\codeF csv}) from github repository.}. For a more conceptual background on NoSQL databases, and NoSQL design methodologies, here are  are related author papers  \cite{asaad2018nosql,asaad2020nosql}

\subsection{NoSQL HBase schema Design}
Confirmed cases and deaths data will be stored respectively in {\code HBase} '{\code confirmed}\_{\code covid19}\_{\code cases}' table, and '{\code deaths}\_{\code covid19}\_{\code cases}' table. Mainly those tables are compliant to JHU CCSE files struture with the first two columns agregation for database  unique key property\footnote{{\codeF time}\_{\codeF series}\_{\codeF  covid19}\_{\codeF confirmed}\_{\codeF global}.{\codeF csv} and {\codeF time}\_{\codeF series}\_{\codeF  covid19}\_{\codeF deaths}\_{\codeF global}.{\codeF csv} under ./{\codeF COVID-19}/{\codeF csse}\_{\codeF covid}\_{\codeF 19}\_{\codeF data}/{\codeF csse}\_{\codeF covid}\_{\codeF 19}\_{\codeF time}\_{\codeF series}/ directory.}.
Each covid-19 row, either for confirmed cases or for deaths, in {\code HBase} will store a country data structured as a composite string primary key ({\code rowid}) constituted from its eventual province/state concatenated with its country name/region and separated with '{\textasciitilde{}}' character. The row then will store all columns values under the same column family '{\code a}' (e.g. '{\code a:lt}' represents latitude, '{\code a:lg}' represents longitude, while remaining dynamic daily dated columns values will be named by convention as '{\code a:d122}' meaning value at January 22nd,  '{\code a:d327}' meaning confirmed cases value of  {\code confirmed}\_{\code covid19}\_{\code cases} table (respectively number of deaths of  {\code deaths}\_{\code covid19}\_{\code cases} table) at March 27nd, etc.

The following {\code HBase} commands retrieve number of confirmed COVID-19 cases, and deaths at March 31st for Morocco (suffix before '~' is empty for all countries) and for British Columbia~Canada (suffix before '~' is not empty for all states) from'{\code confirmed}\_{\code covid19}\_{\code cases}' and '{\code deaths}\_{\code covid19}\_{\code cases}' {\code Hbase} tables.

\begin{minipage}{\textwidth}
\begin{lstlisting}[caption={\code HBase} {\code get} query examples]
get 'confirmed_covid19_cases', '~Morocco', 'a:d331'

get 'deaths_covid19_cases', '~Morocco', 'a:d331'

get 'confirmed_covid19_cases', 'British Columbia~Canada', 'a:d331'

get 'deaths_covid19_cases', 'British Columbia~Canada', 'a:d331'

\end{lstlisting}
\end{minipage}

\subsection{Relational Hive schema Design}
Confirmed cases and deaths data will be respectively represented by two external tables in {\code Hive} '{\code confirmed}\_{\code covid19}\_{\code cases}' table, and '{\code deaths}\_{\code covid19}\_{\code cases}'. Those tables will be relational abstractions mapped ({\it kind of shortcuts pointing}) to their equivalent NoSQL tables in {\code HBase}  (i.e.  non managed Tables - stored physically only in {\code Hbase}).\footnote{In the NoSQL/SQL interoperability between {\codeF HBase} and {\codeF Hive}, the {\codeF Hive}'{\codeF CREATE TABLE} command will create two tables one in {\codeF HBase} and another table in  {\codeF Hive} (the latest is implicitely external)}$^,$\footnote{You should add a new column to {\codeF Hive}/{\codeF HBase} confimed cases, deaths and recovered schemas each day after March 31st, 2020 manually or generate the new schema automatically !!.}$^,$\footnote{SQL {\codeF Hive} {\codeF CREATE TABLE}  commands are may easily be adapted to other relational Big Data store compatible with {\codeF HBase} as Cloudera HDP {\codeF Impala}\cite{impala2020}, IBM Db 2 {\codeF Big SQL}\cite{bigsql2020}, etc.}.

\begin{minipage}{\textwidth}
\begin{lstlisting}[caption=Under HBase remove Confirmed cases Table]
disable 'confirmed_covid19_cases'

drop 'confirmed_covid19_cases'

\end{lstlisting}
\end{minipage}

\begin{minipage}{\textwidth}
\begin{lstlisting}[caption=Under HBase remove Deaths Table]
disable 'deaths_covid19_cases'

drop 'deaths_covid19_cases'
\end{lstlisting}
\end{minipage}

\begin{minipage}{\textwidth}
\begin{lstlisting}[caption=Hive/Hbase Confirmed Cases tables creation]
DROP TABLE confirmed_covid19_cases;

CREATE TABLE confirmed_covid19_cases (
key struct<Province_State : string,Country_Region : string>,
Lat float, 
Long float,
01_22_2020 int, 01_23_2020 int, 01_24_2020 int, 
01_25_2020 int, 01_26_2020 int, 01_27_2020 int,
01_28_2020 int, 01_29_2020 int, 01_30_2020 int,
01_31_2020 int, 02_01_2020 int, 02_02_2020 int,
02_03_2020 int, 02_04_2020 int, 02_05_2020 int,
02_06_2020 int, 02_07_2020 int, 02_08_2020 int,
02_09_2020 int, 02_10_2020 int, 02_11_2020 int,
02_12_2020 int, 02_13_2020 int, 02_14_2020 int,
02_15_2020 int, 02_16_2020 int, 02_17_2020 int,
02_18_2020 int, 02_19_2020 int, 02_20_2020 int,
02_21_2020 int, 02_22_2020 int, 02_23_2020 int,
02_24_2020 int, 02_25_2020 int, 02_26_2020 int,
02_27_2020 int, 02_28_2020 int, 02_29_2020 int,
03_01_2020 int, 03_02_2020 int, 03_03_2020 int,
03_04_2020 int, 03_05_2020 int, 03_06_2020 int,
03_07_2020 int, 03_08_2020 int, 03_09_2020 int,
03_10_2020 int, 03_11_2020 int, 03_12_2020 int,
03_13_2020 int, 03_14_2020 int, 03_15_2020 int,
03_16_2020 int, 03_17_2020 int, 03_18_2020 int,
03_19_2020 int, 03_20_2020 int, 03_21_2020 int,
03_22_2020 int, 03_23_2020 int, 03_24_2020 int,
03_25_2020 int, 03_26_2020 int, 03_27_2020 int,
03_28_2020 int, 03_29_2020 int, 03_30_2020 int,
03_31_2020 int
)
ROW FORMAT DELIMITED
COLLECTION ITEMS TERMINATED BY '\~'
STORED BY 'org.apache.hadoop.hive.hbase.HBaseStorageHandler'
WITH SERDEPROPERTIES (
"hbase.table.name" = "confirmed_covid19_cases",
"hbase.mapred.output.outputtable"="confirmed_covid19_cases",
"hbase.columns.mapping" = ":key,a:lt,a:lg,a:d122,a:d123,
a:d124,a:d125,a:d126,a:d127,a:d128,a:d129,a:d130,a:d131,
a:d201,a:d202,a:d203,a:d204,a:d205,a:d206,a:d207,a:d208,
a:d209,a:d210,a:d211,a:d212,a:d213,a:d214,a:d215,a:d216,
a:d217,a:d218,a:d219,a:d220,a:d221,a:d222,a:d223,a:d224,
a:d225,a:d226,a:d227,a:d228,a:d229,a:d301,a:d302,a:d303,
a:d304,a:d305,a:d306,a:d307,a:d308,a:d309,a:d310,a:d311,
a:d312,a:d313,a:d314,a:d315,a:d316,a:d317,a:d318,a:d319,
a:d320,a:d321,a:d322,a:d323,a:d324,a:d325,a:d326,a:d327,
a:d328,a:d329,a:d330,a:d331",
"hbase.composite.key.factory"="org.apache.hadoop.hive.hbase.SampleHBaseKeyFactory2");

DESCRIBE confirmed_covid19_cases;

\end{lstlisting}
\end{minipage}

\begin{minipage}{\textwidth}
\begin{lstlisting}[caption=Hive/Hbase Deaths table creation]
DROP TABLE deaths_covid19_cases;

CREATE TABLE deaths_covid19_cases (
key struct<Province_State : string,Country_Region : string>,
Lat float, 
Long float,
01_22_2020 int, 01_23_2020 int, 01_24_2020 int, 
01_25_2020 int, 01_26_2020 int, 01_27_2020 int,
01_28_2020 int, 01_29_2020 int, 01_30_2020 int,
01_31_2020 int, 02_01_2020 int, 02_02_2020 int,
02_03_2020 int, 02_04_2020 int, 02_05_2020 int,
02_06_2020 int, 02_07_2020 int, 02_08_2020 int,
02_09_2020 int, 02_10_2020 int, 02_11_2020 int,
02_12_2020 int, 02_13_2020 int, 02_14_2020 int,
02_15_2020 int, 02_16_2020 int, 02_17_2020 int,
02_18_2020 int, 02_19_2020 int, 02_20_2020 int,
02_21_2020 int, 02_22_2020 int, 02_23_2020 int,
02_24_2020 int, 02_25_2020 int, 02_26_2020 int,
02_27_2020 int, 02_28_2020 int, 02_29_2020 int,
03_01_2020 int, 03_02_2020 int, 03_03_2020 int,
03_04_2020 int, 03_05_2020 int, 03_06_2020 int,
03_07_2020 int, 03_08_2020 int, 03_09_2020 int,
03_10_2020 int, 03_11_2020 int, 03_12_2020 int,
03_13_2020 int, 03_14_2020 int, 03_15_2020 int,
03_16_2020 int, 03_17_2020 int, 03_18_2020 int,
03_19_2020 int, 03_20_2020 int, 03_21_2020 int,
03_22_2020 int, 03_23_2020 int, 03_24_2020 int,
03_25_2020 int, 03_26_2020 int, 03_27_2020 int,
03_28_2020 int, 03_29_2020 int, 03_30_2020 int,
03_31_2020 int
)
ROW FORMAT DELIMITED
COLLECTION ITEMS TERMINATED BY '\~'
STORED BY 'org.apache.hadoop.hive.hbase.HBaseStorageHandler'
WITH SERDEPROPERTIES (
"hbase.table.name" = "deaths_covid19_cases",
"hbase.mapred.output.outputtable" = "deaths_covid19_cases",
"hbase.columns.mapping" = ":key,a:lt,a:lg,a:d122,a:d123,
a:d124,a:d125,a:d126,a:d127,a:d128,a:d129,a:d130,a:d131,
a:d201,a:d202,a:d203,a:d204,a:d205,a:d206,a:d207,a:d208,
a:d209,a:d210,a:d211,a:d212,a:d213,a:d214,a:d215,a:d216,
a:d217,a:d218,a:d219,a:d220,a:d221,a:d222,a:d223,a:d224,
a:d225,a:d226,a:d227,a:d228,a:d229,a:d301,a:d302,a:d303,
a:d304,a:d305,a:d306,a:d307,a:d308,a:d309,a:d310,a:d311,
a:d312,a:d313,a:d314,a:d315,a:d316,a:d317,a:d318,a:d319,
a:d320,a:d321,a:d322,a:d323,a:d324,a:d325,a:d326,a:d327,
a:d328,a:d329,a:d330,a:d331",
"hbase.composite.key.factory"="org.apache.hadoop.hive.hbase.SampleHBaseKeyFactory2");

DESCRIBE deaths_covid19_cases;

\end{lstlisting}
\end{minipage}

\subsection{NoSQL HBase Data Loading}
Loading prepared COVID-19 data to {\code HBase} data store is achieved by (i) copying {\code time}\_{\code series}\_{\code  covid19}\_{\code confirmed}\_{\code global-sparse.csv} and \\{\code time}\_{\code series}\_{\code covid19}\_{\code deaths}\_{\code global-sparse.csv} files generated by {\code ingest}\_{\code and}\_{\code clean.sh} script invokations into {\code HDFS} file system, and (ii) then performing bulk looding into {\code HBase} previously created schema.

\begin{minipage}{\textwidth}
\begin{lstlisting}[caption=HBase Feeding with Confirmed Cases data]
#here biadmin is my HDFS user name change it to yours

hadoop fs -rm  /user/biadmin/time_series_covid19_confirmed_global-sparse.csv

hadoop fs -put /home/kbaina/time_series_covid19_confirmed_global-sparse.csv /user/biadmin/

hadoop fs -ls

#here /opt/ibm/biginsights/hbase/bin is my HBase binary directory change it to yours

/opt/ibm/biginsights/hbase/bin/hbase org.apache.hadoop.hbase.mapreduce.ImportTsv -Dimporttsv.separator=',' -Dimporttsv.columns=HBASE_ROW_KEY,a:lt,a:lg,a:d122,a:d123,a:d124,a:d125,a:d126,a:d127,a:d128,a:d129,a:d130,a:d131,a:d201,a:d202,a:d203,a:d204,a:d205,a:d206,a:d207,a:d208,a:d209,a:d210,a:d211,a:d212,a:d213,a:d214,a:d215,a:d216,a:d217,a:d218,a:d219,a:d220,a:d221,a:d222,a:d223,a:d224,a:d225,a:d226,a:d227,a:d228,a:d229,a:d301,a:d302,a:d303,a:d304,a:d305,a:d306,a:d307,a:d308,a:d309,a:d310,a:d311,a:d312,a:d313,a:d314,a:d315,a:d316,a:d317,a:d318,a:d319,a:d320,a:d321,a:d322,a:d323,a:d324,a:d325,a:d326,a:d327,a:d328,a:d329,a:d330,a:d331 -Dimporttsv.skip.bad.lines=true -Dimporttsv.skip.empty.columns=true confirmed_covid19_cases /user/biadmin/time_series_covid19_confirmed_global-sparse.csv

\end{lstlisting}
\end{minipage}

\begin{minipage}{\textwidth}
\begin{lstlisting}[language=sh, caption=HBase Feeding with Deaths data]
#here biadmin is my HDFS user name change it to yours

hadoop fs -rm  /user/biadmin/time_series_covid19_deaths_global-sparse.csv

hadoop fs -put /home/kbaina/time_series_covid19_deaths_global-sparse.csv /user/biadmin/

hadoop fs -ls

#here /opt/ibm/biginsights/hbase/bin is my HBase binary directory change it to yours
/opt/ibm/biginsights/hbase/bin/hbase org.apache.hadoop.hbase.mapreduce.ImportTsv -Dimporttsv.separator=',' -Dimporttsv.columns=HBASE_ROW_KEY,a:lt,a:lg,a:d122,a:d123,a:d124,a:d125,a:d126,a:d127,a:d128,a:d129,a:d130,a:d131,a:d201,a:d202,a:d203,a:d204,a:d205,a:d206,a:d207,a:d208,a:d209,a:d210,a:d211,a:d212,a:d213,a:d214,a:d215,a:d216,a:d217,a:d218,a:d219,a:d220,a:d221,a:d222,a:d223,a:d224,a:d225,a:d226,a:d227,a:d228,a:d229,a:d301,a:d302,a:d303,a:d304,a:d305,a:d306,a:d307,a:d308,a:d309,a:d310,a:d311,a:d312,a:d313,a:d314,a:d315,a:d316,a:d317,a:d318,a:d319,a:d320,a:d321,a:d322,a:d323,a:d324,a:d325,a:d326,a:d327,a:d328,a:d329,a:d330,a:d331 -Dimporttsv.skip.bad.lines=true -Dimporttsv.skip.empty.columns=true deaths_covid19_cases /user/biadmin/time_series_covid19_deaths_global-sparse.csv


\end{lstlisting}
\end{minipage}

\section{Hive SQL/pure NoSQL interoperability and Querying}

Instead of suffering from spreesheats limitations to exploit JHU COVID-19 data with regards to columns number for sorting, or integration of more tables, or versioning different hard coded sheets and workbooks for business users, and instead of coding complex reporting scripts for simple queries for data engineers and data scientists, one may express simple queries both using {\code HBase} and {\code Hive} command line interfaces or through APIs. 

\begin{minipage}{\textwidth}
\begin{lstlisting}[caption=Visualise all confirmed cases and deaths directely from HBase]
scan 'confirmed_covid19_cases'

scan 'deaths_covid19_cases'
\end{lstlisting}
\end{minipage}

\begin{minipage}{\textwidth}
\begin{lstlisting}[caption=HBase queries retrieving  numbers concerning four countries on March 31st 2020]
get 'confirmed_covid19_cases', '~Morocco', 'a:d331'
get 'deaths_covid19_cases', '~Morocco', 'a:d331'

get 'confirmed_covid19_cases', '~Spain', 'a:d331'
get 'deaths_covid19_cases', '~Spain', 'a:d331'

get 'confirmed_covid19_cases', '~France', 'a:d331'
get 'deaths_covid19_cases', '~France', 'a:d331'

get 'confirmed_covid19_cases', '~Germany', 'a:d331'
get 'deaths_covid19_cases', '~Germany', 'a:d331'

\end{lstlisting}
\end{minipage}

\begin{minipage}{\textwidth}
\begin{lstlisting}[caption=Hive query retrieving all confirmed cases data concerning Morocco]
SELECT *
FROM confirmed_covid19_cases
where key.Country_Region='Morocco' ;
\end{lstlisting}
\end{minipage}

\begin{minipage}{\textwidth}
\begin{lstlisting}[caption=Hive Join query retrieving confirmed cases and deaths concerning Morocco on March 31st 2020]
SELECT d.key.Country_Region, c.03_31_2020, d.03_31_2020
FROM confirmed_covid19_cases c
JOIN deaths_covid19_cases d
   ON     c.key.Province_State = d.key.Province_State
      AND c.key.Country_Region=d.key.Country_Region
WHERE c.key.Country_Region ='Morocco' ;
\end{lstlisting}
\end{minipage}

\begin{minipage}{\textwidth}
\begin{lstlisting}[caption=Hive Join query retrieving confirmed cases and deaths concerning four countries on March 31st 2020]
SELECT d.key.Province_State, d.key.Country_Region, c.03_31_2020, d.03_31_2020
FROM confirmed_covid19_cases c
JOIN deaths_covid19_cases d
   ON    c.key.Province_State = d.key.Province_State
     AND c.key.Country_Region=d.key.Country_Region
WHERE c.key.Country_Region in ('Morocco', 'France', 'Spain', 'Germany') ;

\end{lstlisting}
\end{minipage}

\section{Conclusion}
This paper  presents detailed schemas design and data preparation technical {\code HBase}, {\code Hive}, {\code shell} and {\code HDFS} scripts for formatting and storing Johns Hopkins University COVID-19 daily data in {\codeF HBase} NoSQL data store, and enabling {\codeF HiveQL} COVID-19 data querying in a relational {\codeF Hive} SQL-like style. It aims to help scientists and data scientists shortening data preparation phase which is time consuming acording to CRISP-DM life cycle specialists. This work is to be taken as a leveraging bootstrap for specific data preparation phase in COVID-19 analytics Big Data projects aiming for instance  to integrate COVID-19 evolution time series with medical/biology best practices, COVID-19 mutations, scientific papers results, or to study correlations between COVID-19 curves with humidity data, people telco mobilty during countries lockdown phases, or to analyse recurrent COVID-19 contamination causality, or to study similarities with other historical pandemics evolution data like SARS-CoV, MERS-COV, or to compare evolution with spreading information from social networks, etc. The more integration you do on the schema with other data sets (e.g. continents, median age, population, testing numbers, virus contamination rates, etc.), the more features you will have and the more this work will leverage your COVID-19 data experience. Hurry Up, and share you experience for the world scientists.

\section{Appendix : How to download scripts of this paper ?}
To download continuously data engineering models and scripts discussed in this paper, you can access, and clone the author gitlab repository  at \cite{baina2020}.

\section*{Acknowledgment}
Acknowledgement must go to Johns Hopkins University Center for Systems Science and Engineering (JHU CCSE) for keeping up to date world wide COVID-19 data available in a daily frequency.

Acknowledgement must go to The Ministry of National Education, Higher Education, Staff Training, and Scientific Research, Morocco for accepting and supporting my sabbatical leave to do research, and return to ENSIAS refreshed. I also acknowledge my colleagues at ENSIAS maintaining the superb teaching and learning and e-learning culture in the school in my absence especially during COVID-19 crisis.

\bibliographystyle{splncs}
{\small
\bibliography{main.bib}}
}

\end{document}